# Critique of some thermodynamic proofs based on the pump-engine couple

Kamal Bhattacharyya
Department of Chemistry, University of Calcutta, Kolkata-700 009, India
E-mail: pchemkb@yahoo.com

**Abstract:**
   We argue that all the common thermodynamic proofs based on the pump-engine couple possess an inherent circularity and hence are of doubtful value. In certain cases, ways of avoiding them are suggested. In others, the difficulties are discussed. Particularly, we emphasize that a neat proof of Carnot's theorems requires the entropic formulation of the second law.

## 1. Introduction

Sadi Carnot was probably the first to use a pump-engine couple to prove a few theorems named later after him (for historical accounts and relevance, see, *e.g.*, refs. [1-3]). Subsequently, this composite device became quite favourite, so much so that it was freely used by Femi [4, 5], Pauli [6] and Feynman [7], among others, to establish some of Carnot's theorems. Following Fermi's [5] hints, Zemansky [8] and Kubo [9] employed the same couple to show, in addition, the equivalence of the Clausius and Kelvin statements of the second law of thermodynamics. Zemansky and Dittmann omitted a discussion of Carnot's theorems in the sixth edition of their excellent textbook, but included the same later [10]. Pippard [11] and Huang [12] remained faithful to Fermi's earlier work [4] in proving one of Carnot theorems. Thus, virtually all classic texts on thermodynamics relied on the pump-engine couple to confirm one point or more in the context of the second law, notable exceptions being Landsberg [13] and Callen [14].

We shall see here, however, that all such *pump-engine-based* 'proofs' are unsound. In particular, proofs of Carnot's theorems really require the entropic formulation of the second law.

## 2. A summary of standard proofs

All the standard works rest on a general premise. They involve the *reductio ad absurdum* principle. We choose a system *X* that works in a cycle and can act as a heat pump (*P*). Thus, when some work *W* is done on it, it draws heat $Q_1$ from a sink at temperature $T_1$ to release heat $Q_2$ to a source at temperature $T_2$, where $T_2 > T_1$; the temperatures refer to the Kelvin scale. *X* need not be reversible at this stage; only, we *define* its efficiency as $\eta(X) = W/Q_2$. We now take another system *Y* that also works cyclically but acts as a conventional engine (*E*), as depicted in figure 1 with a different set of work and heat terms (small letters). We then couple the two and symbolize the composite device as $Z \equiv X(P)\text{-}Y(E)$ [*X* as a pump and *Y* as an engine]. The proofs (**P**) of several propositions and theorems, all of which we henceforth call theorems (**T**) for the sake of homogeny, now continue as follows. Here, for convenience, we follow the succinct notations of Zemansky [8] ('+' means a true statement, '⊃' means implies).

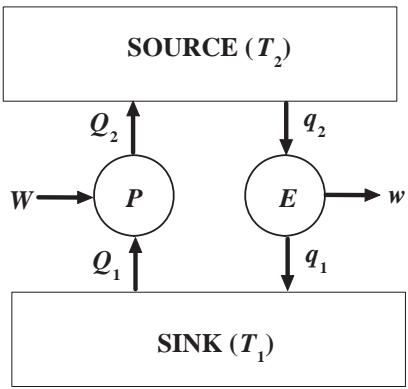

**Figure 1**

    **T1**. The Kelvin (K) and Clausius (C) statements are equivalent.
**P1**: (*a*) Suppose the Kelvin statement is false so that $q_1 = 0$. Choose $W = w$. Device *Z* then transfers $Q_1$ heat from colder to hotter reservoir without any effort, and hence goes against the Clausius statement. Thus, -K ⊃ -C.
    (*b*) Suppose the Clausius statement is false so that $W = 0$. Choose $q_2 = Q_1 + w$. Device *Z* then goes against the Kelvin statement. Thus, -C ⊃ -K.
Coupling (*a*) and (*b*), **T1** is true, *i.e.*, K ≡ C.
    **T2**. No engine is more efficient than a Carnot engine.

**P2a**: Use the Carnot engine as *X* and suppose the converse is true, *i.e.*, $\eta(Y) > \eta(X)$. Accordingly, choose $W = w$ and $Q_2 > q_2$. Then, without any work, *Z* transfers heat $(Q_2 - q_2)$ to the hotter reservoir. Thus, we have $-\mathbf{T2} \supset -\mathbf{C}$. Hence, **T2** can't be false.

**P2b**: Use the Carnot engine as *X* and suppose $\eta(Y) > \eta(X)$. Accordingly, this time we choose $W < w$ but $Q_2 = q_2$. The result for device *Z* then shows $-\mathbf{T2} \supset -\mathbf{K}$. Hence, **T2** is true.

**P2c**: The third proof is due to Fermi [4] and appeared elsewhere too [11, 12]. Here, once again, the same arrangement as above is chosen, *i.e.* the Carnot engine as *X* and the arbitrary engine as *Y*. However, the heat and work terms are chosen arbitrarily. Thus, unlike **P2a** or **P2b**, one takes here $W \neq w$, $Q_2 \neq q_2$. At first sight, this appears the *most* general choice. But, we note that *there is a complete liberty of adjusting the amount of the working substance* in *Y* to change either of the heat terms involved, without affecting its efficiency in anyway. Hence, the complex course of Fermi seems unnecessary. We can safely use his final outcome for device *Z*. This turns out exactly to be **P2b**.

      **T3**. All reversible engines are equally efficient.

**P3**: The following steps are adopted here: (i) Use the Carnot engine as *X* and the other one as *Y*. (ii) Suppose, $\eta(Y) > \eta(X)$. (iii) Use either **P2a**, **P2b** or **P2c** to come to a contradiction and conclude that $\eta(X) \geq \eta(Y)$. (iv) Swap the arrangement to form *Y(P)-X(E)* couple. As both engines are reversible, their $\eta$-values do not change. (v) Suppose now that $\eta(X) > \eta(Y)$. (vi) Employ again either of the proofs for **T2** to face a conflict, Hence, $\eta(Y) \geq \eta(X)$ follows. (vii) Thus, consistency of (iii) and (vi) demands $\eta(X) = \eta(Y)$.

      Some textbooks also establish via the above route that the efficiency of a Carnot engine does not depend on the working substance. Indeed, once **T3** is established, this follows automatically and needs no introduction as a *new* theorem.

## 3. The paradox

      Apparently, in all the proofs above, one *implicitly assumes* that the composite device *Z* is *innocent*. But, this is *not* true and here lies the circularity. In fact, we shall now show that the efficiency of *Y* cannot be greater than that of *X* if *Z*, the *X(P)-Y(E)* couple, is accepted as a *real working device*. Let us carefully note that we have generally two degrees of freedom for engine *Y*, once *X* is kept fixed. If we allow *Y* to extract some heat $q_2$ from the source, do some amount of work $w$ and drain out heat $q_1$ to the sink, then we can specify either $q_2$ and $q_1$, or $q_2$ and $\eta(Y)$ [equivalently, $w$], or $q_1$ and $\eta(Y)$ [equivalently, $w$]. Let us also recall that we can regulate the amount of *the working substance* in *Y* to manipulate *either $q_2$ or $q_1$ or $w$*, keeping $\eta(Y)$ fixed. Thus, there are a number of ways of arriving at the following new theorem:

      **T4**. For an *X(P)-Y(E)* couple to be a real working device, $\eta(Y) \leq \eta(X)$.

**P4a**: Choose, $q_1 = Q_1$ by adjusting the amount of the working substance in *Y*. The demand $\eta(Y) > \eta(X)$ means, respecting energy conservation, that $(1 - Q_1/q_2) > (1 - Q_1/Q_2)$. Therefore, we must have $q_2 > Q_2$. Then, device *Z* takes up some heat from source and converts it into work completely. Thus, $-\mathbf{T4} \supset -\mathbf{K}$. Hence, **T4** is true.

**P4b**: Adjust *Y* such that $w = W$. Then, $\eta(Y) = W/q_2$ and say, if possible, $q_2 < Q_2$. Then the heat given to the sink by *Y* will be $q_2 - W = q_2 - (Q_2 - Q_1)$. This amount is less than $Q_1$. Thus, net result of *Z* will be to drain out heat from a colder to a hotter reservoir without any work. Therefore, $-\mathbf{T4} \supset -\mathbf{C}$ and hence is impossible.

**P4c**: Suppose, by control, $q_2 = Q_2$. Then, $-\mathbf{T4} \supset w > W$. Energy conservation here yields $q_1 < Q_1$. This means, the couple *Z* draws some heat from sink and converts it into work completely. Stated otherwise, $-\mathbf{T4} \supset -\mathbf{K}$, which is impossible.

**P4d**: Suppose, $q_1 > Q_1$ and $q_2 > Q_2$. Then, is it possible that $\eta(Y) > \eta(X)$? Indeed, under a very general condition like this, the following line of reasoning may be used.

      By definition,

$$Q_1 = (1 - \eta(X))Q_2 \tag{1}$$

and
$$q_1 = (1 - \eta(Y))q_2. \tag{2}$$

Take now, if possible,
$$\eta(Y) = \eta(X) + \varepsilon, \varepsilon > 0. \tag{3}$$

It is easy to find then that
$$q_1 - Q_1 = (1 - \eta(X))(q_2 - Q_2) - \varepsilon q_2. \tag{4}$$

Using (3) and (4), we arrive at the following expression for the efficiency of device Z that here works as *an engine*:

$$\eta(Z) = \eta(X) + \varepsilon \frac{q_2}{q_2 - Q_2}. \tag{5}$$

This result is both important and interesting. It shows, if we accept $\eta(Y) > \eta(X)$ initially, it then follows finally that $\eta(Z) > \eta(Y) > \eta(X)$ with Z as a new engine. One can then couple this new engine with pump X to still increase the efficiency of another new composite engine, and so on. Thus, an engine of arbitrarily high efficiency ($\eta \to 1$) becomes possible, surpassing the *Carnot limit* $\eta = (1 - T_1/T_2)$, as no reference to temperature appears anywhere in the argument. This is impossible, we *know* (see later). The contradiction can be avoided only by choosing $\varepsilon < 0$ in which case the inconsequential inequality $\eta(X) > \eta(Y) > \eta(Z)$ would have followed from $\eta(X) > \eta(Y)$. This completes the proof of **T4**.

To be wise after the event, any *arbitrary* couple X(P)-Y(E) is *not* permissible as Z. Such a couple would exist in real world, and hence does not contradict with thermodynamics, *if and only if* $\eta(X) \geq \eta(Y)$. In case both X and Y are reversible, we can also conclude that *both* the X(P)-Y(E) and the Y(P)-X(E) couples are feasible *as* $\eta(X) = \eta(Y)$. This corresponds to choosing $\varepsilon = 0$ in the preceding paragraph.

Having accepted **T4**, we now focus on the 'standard' proofs and point out their limitations. We emphasize that, in the absence of such an a priori analysis of Z as above, contradictions and confusions are quite likely. This is because; an impossible setup is liable to contradict some form of second law at some stage. Here are a few objections:

(1) In **P1**(*a*), the choice $q_1 = 0$, $Q_1 \neq 0$, but $W = w$ contradicts **T4** at the onset. Note that in **P4b**, we didn't use any violation of the Kelvin statement, but still got a final violation of the Clausius statement.

(2) In **P1**(*b*), the starting choice that $\eta(X) = 0$, but $\eta(Y) > 0$, is similarly wrong. We have indeed showed in **P4a** and **P4c** that a violation of the Kelvin statement follows without requiring a concomitant violation of the Clausius one at the start.

(3) In **P2a** or **P2b**, once we take the Carnot pump as X, it is not permissible to choose $\eta(Y) > \eta(X)$. Hence, the route chosen cannot continue. In other words, if we assume a more efficient engine acting as X and the Carnot one as Y, there would be no problem!

(4) The catch in **P3** is obvious. Step (i) requires an *a priori* knowledge that $\eta(X) \geq \eta(Y)$. Still, if everything is all-right up to step (iii), one is permitted to go to step (iv) *iff* $\eta(X) = \eta(Y)$. But, this is exactly what we have set out to prove. In other words, swapping the combination at the very step (iv) assumes what is to be proved. It's just begging the question!

**4. Resolution**

A proof of **T1** is possible as follows. Since +K $\supset q_2 > w > q_1 > 0$, we have the inequality $0 < \eta(Y) < 1$. By **T4**, this yields $\eta(X) \geq \eta(Y)$. It means, $\eta(X) > 0$, *i.e.*, $W > 0$ if $Q_1 > 0$. Thus, +C follows. On the other hand, +C $\supset Q_2 > W > Q_1 > 0$ and so, $\eta(X) < 1$. Hence, by **T4**, $\eta(Y) < 1$, and +K follows.

Consider now **T2**. In view of **T4**, we should be more specific here. If the $X(P)$-$Y(E)$ couple is operative, the condition is $\eta(Y, E) \leq \eta(X, P)$. If the reverse couple is also functional, we would have $\eta(X, E) \leq \eta(Y, P)$. Suppose now that one system is reversible, say $X$. Then, $\eta(X, E) = \eta(X, P)$ follows. The final outcome is

$$\eta(Y,P) \geq \eta(X,P) = \eta(X,E) \geq \eta(Y,E). \tag{6}$$

Note that the inequality at the right of (6) is well-known. But, the argument above does not *prove* **T2** even if $X$ is identified as a Carnot (reversible) engine/pump. A proper running of $Z$ indeed *requires* it. More interesting, however, is the first inequality in (6). Our discussion reveals, it is permissible to have some $Z$ satisfying it. In fine, we do not find a route to prove **T2** or **T3** without importing the concept and definition of entropy and that

$$\Delta S(U) \geq 0, \tag{7}$$

where '$U$' refers to the universe.

Nevertheless, device $Z$ does identify the *special* status of a reversible (or, Carnot) engine/pump in the following way. Consider again the $X(P)$-$Y(E)$ couple with $\eta(Y, E) \leq \eta(X, P)$, in accordance with **T4**. We thus take

$$\eta(X) = \eta(Y) + \varepsilon, \varepsilon > 0. \tag{8}$$

We also choose

$$Q_1 = \kappa(q_1 - \varepsilon q_2), Q_2 = \kappa q_2, 0 < \kappa < 1, \tag{9}$$

such that $q_1 > Q_1$ and $q_2 > Q_2$. This ensures $Z$ to act as an engine. One can easily check then that

$$\eta(Z,E) = \eta(Y,E) - \frac{\kappa \varepsilon}{1-\kappa}. \tag{10}$$

If, on the other hand, we had chosen, along with (8),

$$q_1 = \kappa(Q_1 + \varepsilon Q_2), q_2 = \kappa Q_2, 0 < \kappa < 1, \tag{11}$$

instead of (9), so as to satisfy $q_1 < Q_1$ and $q_2 < Q_2$, device $Z$ would have served as a pump. In that case we would likewise find

$$\eta(Z,P) = \eta(X,P) + \frac{\kappa \varepsilon}{1-\kappa}. \tag{12}$$

One may now ask: 'when would $\eta(Z, E)$ attain a maximum value?' As (10) shows, this happens in the limit $\varepsilon \to 0$, and then we also have $\eta(Z, E) = \eta(Y, E) = \eta(X, P)$. This means, we can invert the $X(P)$-$Y(E)$ couple as well. A similar question 'when would $\eta(Z, P)$ attain a maximum value?' is not significant. A better pump should have lesser $\eta$, by the definition employed here. Hence, on the basis of (12), we seek a situation to obtain a minimum $\eta(Z, P)$. This again occurs as $\varepsilon \to 0$, implying $\eta(Z, P) = \eta(Y, E) = \eta(X, P)$. Only with *reversible cycles*, either of the two possibilities can be realized. In passing, we also notice another character of reversible cycles. They only can form a couple as above that may be run to act as *a perfect insulator*. Note that we avoided the use of 'coefficient of performance' of a pump to keep things simple.

**5. Discussion**

To prove **T2**, we need to consider two engines, not the pump-engine couple. Since any engine works cyclically, its entropy change over a full cycle is zero. Hence, $\Delta S(U)$ would involve only the changes of the reservoirs. Referring to Figure 1, we find

$$\Delta S(U) = \frac{q_1}{T_1} - \frac{q_2}{T_2}. \tag{13}$$

For a reversible (Carnot) engine $C$, i.e., with $Y$ as $C$, one finds from (7) the minimum possible value for (13), which is zero. Indeed, the equality in (7) *defines* reversibility and thus we have the maximum efficiency of a Carnot engine as $\eta(C, E) = 1 - q_1/q_2 = 1 - T_1/T_2$. Incidentally, this now shows the adequacy of proof **P4d**. For any other engine $Y$ involving irreversible steps, the same change of state during interaction with the source (sink) requires less absorption (more rejection) of heat. Thus, $q_2' < q_2$ and $q_1' > q_1$, so that one obtains $\Delta S(U) > 0$. In other words, $\eta(Y, E) = 1 - q_1'/q_2' \leq \eta(C, E)$ follows. Similarly, $\eta(X, P) = 1 - Q_1'/Q_2' \geq \eta(C, P)$ is ensured via a similar logic. Since (7) is concerned only with reversibility or irreversibility, and has no special reference to the Carnot engine or pump, **T3** follows immediately. The nature of the *working substance* also plays no role here. One can easily justify **T4** via (7) as well.

It is quite possible that Carnot could not discover a *logical fallacy* in the construction of $Z$ because, a theorem like **T4** needs either the Kelvin (see **P4***a* or **P4***c*) or the Clausius (see **P4***b*) statement, none of which was available at his time. He could not choose path **P4***d* either, as it *requires* that his own engine possesses the maximum efficiency, which was his intention to prove. However, the genius of Carnot in planning the couple $Z$ may be appreciated from the following observations: (i) his engine part $Y$ gave rise to the Kelvin statement, (ii) the pump $X$ led to the Clausius statement and (iii) the special status of reversible cycles nicely follows from $Z$ (see, e.g., the discussion accompanying (8) and below).

**6. Conclusion**

In summary, our major finding is that, Carnot's theorems cannot be proved without having recourse to the proper form of the second law involving entropy. The existing proofs based on the pump-engine couple are erroneous. It is really startling to note that the inherent circularity escaped attention of even the great minds [4 - 7]. Indeed, this may be a good subject for the historians of science. We have, however, also noticed some special properties of Carnot's couple, especially in the context of reversibility.

We should next remark that the impacts of Carnot's works and the second law have proved to be enormous till today (see, e.g. [15-17], particularly [16]). An interesting early work [18] revealed that Nernst theorem and its conclusions follow as well from the second law. It is now also accepted [19] that the concept of a reversible Carnot machine is *not* essential in defining entropy. This is pretty satisfying since we have found here one important blockade towards the traditional approach to entropy. Mention may be also be made of another recent work [20] that showed, a violation of the Clausius inequality is possible without affecting the second law in the form of (7). This stands as another obstruction. Indeed, such works point only to the weaknesses of the traditional historical development. One way to avoid the route is to cling to some sort of axiomatic approach [13, 14]. A number of other such courses have appeared from time to time, starting with Caratheodory's work. This field, however, is again quite vast and open (for a recent exposition, see [21], with references to earlier works). For example, it has been noted that the functional properties of the entropy gives rise to *six* possible types of thermodynamics [22]. A neat, universal route probably seems demanding.

In fine, we have used throughout $\eta$ as the ratio of work done and heat exchanged with the source. In practical terms, this quantity measures the efficiency of an engine, but inefficiency of a pump. However, that does not invalidate the logic as long as we employ $\eta$ properly [see, *e.g.*, the discussion below (12)]. We also consciously used $T$ to mean temperatures in the Kelvin scale since the beginning. This is because, it appears from the present work that one cannot possibly have an independent thermodynamic temperature scale.

**References**
1. Erlichson H 1999 *Eur. J. Phys.* **20** 183.
2. Uffink J 2001 Bluff your way in the Second Law of Thermodynamics, arXiv:cond-mat/0005327 v2.
3. Truesdell C A III and Bharatha S 1977 Concepts and Logic of Classical Thermodynamics as a Theory of Heat Engines (Springer-Verlag, Heidelberg).
4. Fermi E 1956 Thermodynamics (Dover, New York).
5. Fermi E 1966 Notes on Thermodynamics and Statistics (University of Chicago Press, Chicago).
6. Pauli W 2000 Thermodynamics and the Kinetic Theory of Gases (Dover, New York).
7. Feynman R P, Leighton R B and Sands M 1963 The Feynman Lectures on Physics, Vol. 1 (Addison-Wesley, Reading, Massachusetts).
8. Zemansky M W 1957 Heat and Thermodynamics (McGraw-Hill, Kogakusha, 4$^{th}$ Ed.).
9. Kubo R 1968 Thermodynamics (North-Holland, Amsterdam).
10. Zemansky M W and Dittman R H 1997 Heat and Thermodynamics (McGraw-Hill, Singapore, 7$^{th}$ Ed.).
11. Pippard A B 1960 Elements of Classical Thermodynamics (University Press, Cambridge).
12. Huang K 1963 Statistical Mechanics (John Wiley, New York).
13. Landsberg P T 1961 Thermodynamics with Quantum-Statistical Illustrations (Interscience, New York).
14. Callen H B 1985 Thermodynamics and an Introduction to Thermostatistics (John Wiley, New York).
15. Nikulov A and Sheehan D 2004 *Entropy* **6** 1.
16. Hannay J H 2006 *Am. J. Phys.* **74** 1.
17. Abou Salem W K and Fröhlich J 2007 *J. Stat. Phys.* **126** 1045.
18. Yan Z and Chen J 1988 *J. Phys. A: Math. Gen.* **21** L707.
19. Lieb E H and Yngvason J 1999 *Phys. Rep.* **310** 1
    Lieb E H and Yngvason J 2000 A Fresh Look at Entropy and the Second Law of Thermodynamics, Vienna, Preprint ESI 858.
20. Gavrilov A 2006 The Clausius Inequality does not follow from the Second Law of Thermodynamics, arXiv:physics/0611044 v1, 2006
    Gavrilov A 2007 *On the Clausius Theorem*, arXiv:physics.gen-ph/0077.3871 v1.
21. Pogliani L and Berberan-Santos M N 2000 *J. Math. Chem.* **28** 313.
22. Landsberg P T 1999 *Braz. J. Phys*, 29 46.